\documentclass[aps,pra,twocolumn,showpacs,preprintnumbers,amsmath,amssymb]{revtex4}
 \usepackage{graphicx}
 \usepackage{dcolumn}
 \usepackage{bbm}

\begin{document}
 \bibliographystyle{apsrev}

 \preprint{APS/123-QED}

 \title{Dissipation in gases trapped in time--dependent external potentials}% Force line breaks with \\

 \author{M. Wenin}
 \email{markus.wenin@recendt.at}
  %\altaffiliation[Electronic Adress Markus.Wenin@uni-graz.at ]{.}%Lines break automatically or can be forced with \\

 \affiliation{%
\\ Research Center for Non Destructive Testing (RECENDT), Science Park 2, Altenberger Strasse 69, 4040
Linz, Austria
 }%
 \date{\today}% It is always \today, today,

\begin{abstract}
We investigate an ideal gas in a time--dependent external trapping
potential. We use the Boltzmann equation with the relaxation time
ansatz to explore the time--dependent energy of an adiabatically
isolated system. In particular we are interested on the
dissipation during a potential change along a given protocol with
finite velocity. The role of the relaxation less evolution as a
limiting case is studied: starting from an equilibrium
distribution and for small times the energy of the gas with
relaxation is always smaller than that of the relaxation--less.
This means that relaxation processes show an ambivalent behavior:
on the one hand entropy production, but on the other hand
reduction of dissipation by driving back the system into
equilibrium.
   \\
\end{abstract}
%%Pacs sind bereits richtig
 \pacs{47.45.Ab, 51.10.+y, 05.20.Dd,47.70.Nd, 05.70.Ln}
 \maketitle
 \section{\label{sec:level1} Introduction}
Dissipation of energy in gases or liquids due to a moving body is
at the core of physical theories such as hydrodynamics,
statistical physics or kinetic theory. Some interesting new
results at a fundamental level, regarding the connection of
equilibrium and non--equilibrium free energy differences, as the
Jarzynski equation, or the relation between dissipation and the
asymmetry of time--reversal processes are
known\cite{Jarzynski,Crooks,Seifert,Broeck,Parrondo}. These
results have been extended from classical to quantum mechanical
systems \cite{Mukamel,Nolte,Gaspard}. In this work we investigate
the dissipation process in gases using the Boltzmann equation (BE)
with a time--dependent external potential. The main motivation for
considering this problem was the ambivalent behavior of elastic
collisions during the potential deformation: on the one hand
collisions do not change the energy of the colliding particles,
and this is also true for a time--dependent potential, since the
collisions are described as pointlike in phase space. On the other
hand it is immediately clear that the energy of the gas in a
time--dependent potential must depend on the collisions during the
potential change. It makes a difference in the distribution
function if in the BE collisions are present or not. The first
impulse to investigate such problems using kinetic theory was the
study of Bose--Einstein condensation in trapped gases, where one
can reach the condensation of a Bose gas by
changing the trapping potential \cite{Stamper-Kurn,Stringari}.\\
The paper is organized as follows: Sec. ~\ref{sect2} gives a short
overview of the theoretical elements and formulas used to derive
the main results represented in Sec. ~\ref{sect3}.
Sec.~\ref{sect4} gives a numerical example and Sec.~\ref{sect5}
conclusion and outlook. The appendix contains a proof of the
H--Theorem for time--dependent potentials and the relaxation time
ansatz in the BE.
 \section{General approach: kinetic theory}\label{sect2}
First we present some general equations of kinetic theory used in
this paper. The central quantity to describe the system is the
time--dependent one particle distribution function $f({\textbf
p},{\textbf x},t)$. The distribution function depends on the one
particle phase space coordinates $\{{\textbf p},{\textbf x}\}$
(momentum and space coordinates, sometimes suppressed to simplify
notation) and time $t$ (also suppressed sometimes). $f$ is
normalized to the number of particles $N$,
\begin{equation}\label{f_gN0}
\int f({\textbf p},{\textbf x},t)\mathrm{d}\Gamma=N~.
\end{equation}
Here $\mathrm{d}\Gamma =\mathrm{d}^3 x \mathrm{d}^3 p/(2\pi)^3$
denotes the volume element in the 6 dimensional phase space (we
omit always Planck's constant $\hbar$ and the Boltzmann constant
$k_B$).
 \subsection{Kinetic equation}
We consider a gas in a time--dependent external potential
$U({\textbf x},t)$. The time evolution of the distribution
function is given by the Boltzmann equation (see for example
\cite{LandauX})
\begin{equation}\label{BE}
\frac{\partial f}{\partial t}+\frac{{\textbf
p}}{m}\cdot\nabla_{\textbf x} f-\nabla_{\textbf x} U({\textbf
x},t)\cdot \nabla_{\textbf p} f=\left(\frac{\partial f}{\partial
t}\right)_{coll}~.
\end{equation}
Here $(\partial f/\partial t)_{coll}$ denotes the collision
integral, which we approximate by the relaxation time ansatz,
\begin{equation}\label{coll}
\left(\frac{\partial f}{\partial
t}\right)_{coll}=-\frac{f-f_g}{\tau}~.
\end{equation}
Because the external potential is time--dependent the equilibrium
distribution $f_g({\textbf p},{\textbf x},t)$ becomes
time--dependent as well,
\begin{equation}\label{f_g}
f_g({\textbf p},{\textbf x},t)=\exp\{-\beta(t)[H({\textbf
p},{\textbf x},t)-\mu(t)]\}~.
\end{equation}
Here $\mu(t)$ is the time--dependent chemical potential,
$\beta(t)=1/T(t)$ corresponds to the time--dependent inverse
temperature $T(t)$ and
\begin{equation}
H({\textbf p},{\textbf x},t)=\frac{{\textbf p}^2}{2m}+U({\textbf
x},t)~,
\end{equation}
is the one particle Hamiltonian for particles with mass $m$ and
without inner degrees of freedom. The equilibrium distribution
Eq.~\eqref{f_g} does not contain a macroscopic momentum or angular
momentum because we restrict ourselves to spherical symmetric
potentials to make the discussion transparent. The main result
Eq.~\eqref{res} is not affected by this restriction, as the
derivations show, since the explicit structure of $f_g$ does not
enter. So let us assume that the number of particles and the
energy for a fixed time are the only conserved quantities, which
allow the determination of $\beta(t)$ and $\mu(t)$ in
Eq.~\eqref{f_g}. In particular we have the relation
\begin{equation}\label{f_gN}
\int f_g({\textbf p},{\textbf x},t)\mathrm{d}\Gamma=\int
f({\textbf p},{\textbf x},t)\mathrm{d}\Gamma~,
\end{equation}
ensuring conservation of the number of particles. The conservation
of energy $E(t)$ for a fixed time is expressed as
\begin{eqnarray}\label{f_gE}
E(t)=\int H({\textbf p},{\textbf x},t) f_g({\textbf p},{\textbf
x},t)\mathrm{d}\Gamma=\nonumber{}\\ \int H({\textbf p},{\textbf
x},t) f({\textbf p},{\textbf x},t)\mathrm{d}\Gamma~.
\end{eqnarray}
Eq.~\eqref{BE} with an initial condition (in  our case a
Maxwell--Boltzmann distribution) and the two conditions
Eq.~\eqref{f_gN} and Eq.~\eqref{f_gE} are the basis for our
analytical and numerical studies in Sec.~\ref{sect4}. We remark
that Eq.~\eqref{BE} due to Eq.~\eqref{f_gN} and Eq.~\eqref{f_gE}
becomes non linear, in contrast to the Focker--Planck equation,
which is a linear evolution equation.
\subsection{Entropy}
A central quantity is the nonequilibrium entropy of the gas. For a
given one particle distribution function it is computed as
\cite{LandauX}
\begin{equation}\label{S}
S(t)=-\int f({\textbf p},{\textbf x},t)\ln[f({\textbf p},{\textbf
x},t)]\mathrm{d}\Gamma~.
\end{equation}
One can proof the validity of the H--Theorem, $\dot{S}(t)=-\int
\ln(f)(\partial f/\partial t)_{coll}\mathrm{d}\Gamma \geq 0$ with
the relaxation time ansatz and a time--dependent external
potential also. For details see the appendix App.~\ref{app1}. Here
we need an additional expression for $S(t)$ for gases not far from
equilibrium. We split the distribution function as $f=f_g+\delta
f_g$, where $\delta f_g$ denotes the small deviation of $f$ from
the equilibrium. Expansion of the logarithm in Eq.~\eqref{S} up to
second order and some lines of algebraic simplifications yields
\begin{equation}\label{Sapp}
S(t) = S_{eq}(t)-\int\frac{\delta f_g^2}{2 f_g}\mathrm{d}\Gamma~.
\end{equation}
Here $S_{eq}(t)=\beta(t) E(t) - \beta(t)\mu(t) N$ is the
time--dependent equilibrium entropy, reached when $\delta f_g=0$
\footnote{To simplify the notation we write $S_{eq}(t)$ instead of
$S_{eq}(\omega(t))$, where $\omega(t)$ is the time--dependent
parameter in the Hamiltonian.}. Due to relaxation $\delta f_g(t)$
becomes smaller and smaller and at the end (for a fixed potential)
$S(t\gg \tau)=S_{eq}$.

\section{Distribution function: Perturbation theory}\label{sect3}
To proceed we have to compute the distribution function. Because
the analytical integration of the kinetic equation for
time--dependent potentials (even with the simple relaxation time
ansatz) in general is not possible we look for a perturbation
theory approach. There are two interesting limiting cases as
starting points for this attempt: an infinitely slow potential
change, corresponding to the adiabatic time evolution of the
distribution function, and the relaxation less time evolution
[$\tau\rightarrow\infty$ in Eq.~\eqref{BE}], corresponding to the
Liouville evolution (denoted as $f_L$), respectively. We choose
here the second way, where the system has a large relaxation time,
compared with the time $t$, for which we are interested, so that
$t/\tau \ll 1$. An analytical integration of the time--dependent
collision less BE is still impossible in general, but here we do
not need the solution explicitly.
\subsection{First order expression}
The lowest (zeroth) order of such an expansion is the
time--dependent Liouville function $f_L$. We split the true
distribution function into two parts,
\begin{equation}\label{fl}
f = f_L+\delta f_L
\end{equation}
where $f_L({\textbf p},{\textbf x},t)$ is the solution of the
collision less BE  and $\delta f_L$ is the deviation from the
actual distribution. If we enter with this ansatz into
Eq.~\eqref{BE}, we get for the first order correction
\begin{equation}\label{df}
\delta f^{(1)}_L({\textbf p},{\textbf
x},t)=-\frac{1}{\tau}\int_{0}^{t}\mathrm{d}t'[f_L({\textbf
p},{\textbf x},t')-f^{(0)}_g({\textbf p},{\textbf x},t')]~.
\end{equation}
This first order expression is sufficient as long as the ratio
$t/\tau$ is small. The equilibrium distribution $f^{(0)}_g$ is
computed using $f_L$ on the right hand side of the conditions
Eq.~\eqref{f_gN}, Eq.~\eqref{f_gE}. We do not require the
computation of $f_g^{(0)}$, but we know that it does not depend on
the relaxation time $\tau$.
\subsection{Energy}
The distribution function Eq.~\eqref{fl}  with Eq.~\eqref{df}
allows us to clarify the influence of $\tau$ on the dissipated
energy. We consider the begin of a potential change, where the
distribution function deviates weakly from equilibrium. The
starting point is Eq.~\eqref{Sapp}, where we replace $S(t)$, using
Eq.~\eqref{dotSapp}, by
\begin{equation}\label{S_st}
S(t)=S(0)+\frac{t}{\tau}\int \frac{\delta
f_g^2}{2f_g^{(0)}}\mathrm{d}\Gamma~.
\end{equation}
Now we use $f$ from Eq.~\eqref{fl}, where we approximate the time
integral in Eq.~\eqref{df} by a trapezoidal rule. After some
simplifications we obtain up to the order $(t/\tau)^2$,
\begin{equation}
S_{eq}(t)=S(0)+\left[1-\left(\frac{t}{\tau}\right)^2\right]\int
\frac{(f_L-f_g^{(0)})^2}{2 f_g^{(0)}}\mathrm{d}\Gamma~.
\end{equation}
The dissipated energy (heat) is given by $T(t)[S_{eq}(t)-S(0)]$.
For small times the choice $T(t)=T_{ad}(t)$ is reasonable and we
get\footnote{$T_{ad}(t)$ means the temperature change for an
adiabatic process. It is also possible to set here the initial
temperature $T_0$ or a weighted sum of both.}
\begin{equation}\label{E_st}
E(t)=E_{ad}(t)+T_{ad}(t)\left[1-\left(\frac{t}{\tau}\right)^2\right]\int
\frac{(f_L-f_g^{(0)})^2}{2 f_g^{(0)}}\mathrm{d}\Gamma~.
\end{equation}
For $\tau\rightarrow \infty$ we obtain from this expression
\begin{equation}
E_L(t):=\int H f_L \mathrm{d}\Gamma=E_{ad}(t)+T_{ad}(t)\int\frac{
(f_L-f_g^{(0)})^2}{2 f_g^{(0)}} \mathrm{d}\Gamma~.
\end{equation}
and further (for $t/\tau \ll 1$)
\begin{equation}\label{res}
E(t)\leq E_L(t)~.
\end{equation}
This result shows that the time--dependent energy $E_L$ represents
an upper bound of the energy of the gas, independent of the
characteristics of the potential. Relaxation favors minimization
of dissipation. This is remarkable because the Liouville evolution
itself conserves entropy $S(t)=-\int f_L \ln(f_L)\mathrm{d}\Gamma
=S(0)$. For larger times Eq.~\eqref{res} can be violated, however
for slow potential changes it must fulfilled, because the true
distribution functions $f$ and $f_g$ approaches the adiabatic
distribution function $f_{ad}$. Because the dissipated energy is
given by $E_{diss} = E - E_{ad}$ one obtains $E_{diss}\leq
E_L-E_{ad}$.
\section{Numerical example}\label{sect4}
For a model potential we consider a one dimensional harmonic
oscillator with time--dependent frequency, as widely used to
investigate dissipation of Brownian particles and the Jarzynski
equation \cite{Speck,Nolte}. An increase/decrease of the frequency
leads to a compression/expansion of the gas. We integrate
numerically the BE with the relaxation time ansatz and monitor the
energy (and entropy) of the gas during the potential change. The
numerical method is based on a discretization of the phase space
into equal sized cells and a numerical integration of the
resulting ordinary differential equations. To obtain the
time--dependent equilibrium distribution $f_g$, we compute for
each time step the integrals Eq.~\eqref{f_gN} and Eq.~\eqref{f_gE}
and solve numerically the resulting equations for $\beta$ and
$\mu$ (where the values for the quasistatic potential change are
used as a starting guess). This procedure runs well and there are
no numerical problems which we should report. To be sure that
energy conservation (not only number of particles) for a fixed
potential is fulfilled, the used numerical limits of the phase
space must be large enough. The external potential has the form
\begin{equation}\label{Uex}
U(x,t)=\frac{1}{2}m \omega(t)^2x^2~,\hspace{5mm}
\omega(t)=\omega_0+\Delta\omega\sin(2\pi t/t_f)~.
\end{equation}
Here $t_f$ characterize the switching time and likewise denotes
the final integration time. We set $t_f=1$ as our time scale. We
choose $\omega_0 t_f=1$ and $|\Delta \omega/\omega_0|=0.5$, so we
are far away from adiabatic conditions. The mean free fly time of
a particle is on the order $\sim 1/(\omega_0 N)$, where we have
choosen $N=1250$ (and the ratio initial temperature to initial
frequency $T_0/\omega_0=200$). For Bose gases at low temperatures
and densities as H and Na in magneto--optical traps the relaxation
time differs dramatically due to the different $s$-wave scattering
cross sections (by a factor of $\sim10^3$)\cite{Walraven}. For an
adiabatic potential change the temperature of the gas is given by
$T_{ad}/T_0=\omega/\omega_0$ and the energy
$E_{ad}/E_0=\omega/\omega_0$, respectively\footnote{For
instantaneous switching $\omega_0\rightarrow \omega$ the energy
changes as $E_\infty/E_0=\frac{1}{2}[1+(\omega/\omega_0)^2]$.}.
\begin{figure}
\begin{center}
\includegraphics[height=60 mm, angle=0]{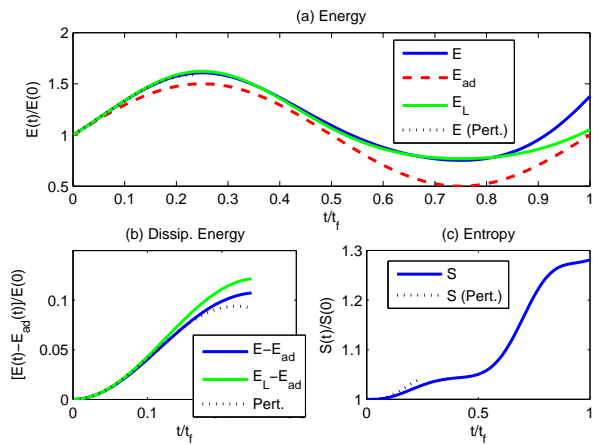}
\caption{\label{fig1} (color online) (a) Energy of the gas during
the compression and expansion. The frequency ratio is
$\Delta\omega/\omega_0=0.5$. The green line corresponds to the
Liouville evolution ($E_L$) and the blue line to a evolution with
finite relaxation time ($\tau/t_f=0.5$). The red dotted line is
the adiabatic limit and in black--dotted it is shown the result
from the perturbation theory. (b) shows the dissipated energy for
a time regime where Eq.~\eqref{E_st} is valid. In (c) we have
plotted the entropy of the gas and the approximation according
Eq.~\eqref{S_st}. For small times $E_L$ is the upper bound for the
energy. }
\end{center}
\end{figure}
\begin{figure}
\begin{center}
\includegraphics[height=60 mm, angle=0]{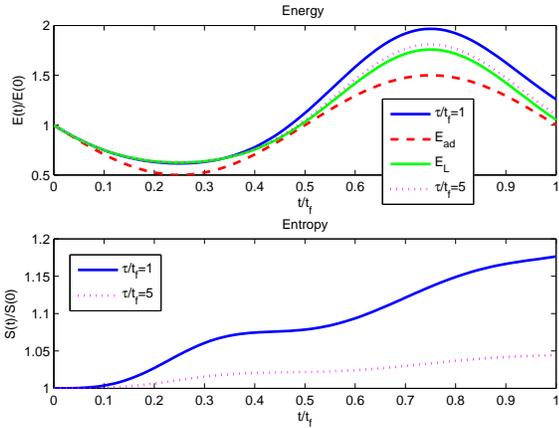}
\caption{\label{fig2}(color online) Energy of the gas during the
expansion and compression cycle, $\Delta\omega/\omega_0=-0.5$.
Although it is difficult to see the energy corresponding to the
Liouville evolution $E_L$ for small times is always greater than
all others, in accordance to Eq.~\eqref{res}.}
\end{center}
\end{figure}
\begin{figure}
\begin{center}
\includegraphics[height=70 mm, angle=0]{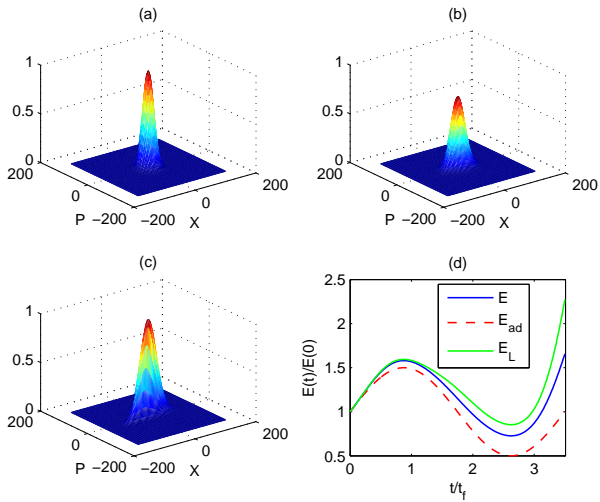}
\caption{\label{fig3}(color online) The same potential deformation
as in Fig.~\ref{fig1} but 3.5 times slower as there (with
$\tau=t_f$). In (a) it is shown the initial Maxwell--Boltzmann
distribution in phase space [in units $(P,X)=(p/\sqrt{m
T_0},x\sqrt{m \omega_0^2/ T_0})$], (b) shows the partially relaxed
distribution at $t=t_f$ and (c) shows the Liouville distribution
$f_L(t_f)$. In (d) we have plotted again the energy of the gas
(blue line) to demonstrate the validity of Eq.~\eqref{res} for the
entire time.}
\end{center}
\end{figure}
In Fig.~\ref{fig1} we show the energy/entropy for one compression
-- expansion chicle as a function of time for a gas with specified
relaxation time. The lines for the energy reflect the
time--dependent frequency, whereas the entropy increase as
required from the H--theorem. We have also plotted the
perturbation theory expressions Eq.~\eqref{E_st} and
Eq.~\eqref{S_st} to demonstrate the result Eq.~\eqref{res}.
Fig.~\ref{fig2} demonstrates the same behavior of the gas for the
opposite chicle, where first the frequency $\omega$ decrease and
hereafter increase (expansion and compression). Here we have
plotted the time--dependent energy of two gases with different
relaxation times and additional $E_{ad}(t)$ and $E_L(t)$. For
small times $E_L(t)$ is an upper bound of the energy.
Fig.~\ref{fig3} shows the distribution function, obtained by the
numerical integration. The initial Maxwell--Boltzmanm distribution
(a) and the collision less evolved function (c), as well as an
example with finite relaxation time (b) the are plotted. In this
case (slower potential change as in the previous example) the
energy of the $\tau\neq 0$ system is smaller as $E_L$ for the
entire switching time.

\section{Conclusion and outlook}\label{sect5}
In this paper we have discussed energy dissipation in gases in
time--dependent trapping potentials. Using the Boltzmann equation
with the relaxation time ansatz we have shown that the dissipated
energy depends on the relaxation time. In particular we have
demonstrated using perturbation theory, that for small times
(compared with the relaxation time) the relaxation less time
evolution always leads to a higher energy than that with finite
relaxation time. On this time scales relaxation decreases
dissipation. The derivation shows that this statement is fairly
general: it does not depend on the shape of the external
potential. It would be interesting to make some numerical studies
with the full collision integral of the BE to check if the results
of this paper are still valid \cite{Bao,Lobo}.

\appendix
\section{H--Theorem for time--dependent potentials and the relaxation time ansatz}
\label{app1} The time derivative of the entropy in the relaxation
time ansatz is equal
\begin{equation}\label{sdot}
\dot{S}(t)=\int \ln(f) \frac{f-f_g}{\tau}\mathrm{d}\Gamma~.
\end{equation}
If we set inside the logarithm $f=f_g+\delta f_g$ we obtain
\begin{eqnarray}
\int \ln(f) \frac{f-f_g}{\tau}\mathrm{d}\Gamma=\frac{1}{\tau}\int
f \ln(f) \mathrm{d}\Gamma-{}\nonumber\\ \frac{1}{\tau}\int
f_g\ln\left[f_g\left(1+\frac{\delta f_g}{f_g}\right)\right]
\mathrm{d}\Gamma~.
\end{eqnarray}
Now we split the argument of the logarithm in to two parts,
\begin{equation}\label{Sz}
\dot{S}(t)=\frac{S_{eq}(t)-S(t)}{\tau}-\frac{1}{\tau}\int
f_g\ln\left[1+\frac{\delta f_g}{f_g}\right] \mathrm{d}\Gamma~,
\end{equation}
where we have defined the equilibrium entropy $S_{eq}(t)=-\int
f_g\ln(f_g)\mathrm{d}\Gamma$. Because $f_g$ is the equilibrium
distribution, corresponding to the same number of particles and
energy as $f$, it follows $S_{eq}(t)-S(t)\geq 0$. The last
integral in Eq.~\eqref{Sz} is positive for all $t$. If $x>0$ is a
real number, then $\ln(x)\leq x-1$ is always true. Using this for
the integral above we obtain ($f_g\geq 0$),
\begin{equation}
\int f_g\ln\left[1+\frac{\delta f_g}{f_g}\right]
\mathrm{d}\Gamma\leq \int \delta f_g \mathrm{d}\Gamma=0~.
\end{equation}
With the $-$ sign in front of the integral in Eq.~\eqref{Sz} we
conclude that $\dot{S}\geq 0$.\\
In the text we use an expression for $\dot{S}$ for small $\delta
f_g$. One can obtain this expression by insertion of $f=f_g+\delta
f_g$ into the logarithm in Eq.~\eqref{sdot} and a series
expansion,
\begin{equation}\label{dotSapp}
\dot{S}(t)\approx \frac{1}{\tau}\int \frac{\delta f_g^2}{f_g}
\mathrm{d}\Gamma~.
\end{equation}

% \bibliography{referenzliste}
 \end{document}